\documentclass[prb,twocolumn,groupedaddress,superscriptaddress]{revtex4}
\usepackage{eurosym}
\usepackage{graphicx}
\usepackage{amssymb}
\usepackage{latexsym}
\usepackage[usenames]{color}
\usepackage{float}

\begin{document}

\title{Energy emanating from the molecular nanomagnet Fe$_{8}$ revisited}
\author{Tom Leviant}
\affiliation{Department of Physics, Technion - Israel Institute of Technology, Haifa,
32000, Israel}
\author{Shaul Hanany}
\affiliation{School of Physics and Astronomy, University of Minnesota/Twin Cities, 116 Church Street SE Minneapolis, MN 55455}

\author{Yuri Myasoedov }
\affiliation{Department of Condensed Matter Physics, The Weizmann Institute of Science, 76100 Rehovot, Israel}
\author{Amit Keren}
\affiliation{Department of Physics, Technion - Israel Institute of Technology, Haifa,
32000, Israel}
\date{\today }

\begin{abstract}

In the molecular nanomagnet Fe$_{8}$ tunneling can occur from a metastable state to an excited state followed by a transition to the ground state. This transition is accompanied
by an energy release of $115.6$~GHz. We constructed an experimental setup to measure whether
this energy is released in the form of thermal or electromagnetic energy. Contrary
to a previous publication we find no evidence for release of electromagnetic radiation.
Our results for transitions between the first and second excited states to the ground
state are consistent with a release of only thermal energy. This energy release extends for a longer time for the second excited state than for the first excited state.
\end{abstract}

\maketitle

While investigating the Fe$_{8}$ mononuclear magnet Shafir and Keren \cite%
{Oren} made a serendipitous observation; tunneling events where accompanied by
a jump in the temperature of a thermometer placed far from the sample and
attached directly to the mixing chamber of a dilution refrigerator (DR).
When the line of site between the thermometer and sample was blocked, the
tunneling signal remained, but the temperature jumps disappeared. This led
to the conclusion that energy bursts accompany the tunneling event
and arrive at the thermometer in the form of electromagnetic radiation. In
order to block the line of site the DR had to warm up and cool down again.
Therefore, the test experiment was not done simultaneously with main
experiment. Here we revisit the same phenomena, but with an experimental
setup designed to detect photons in the microwave range, and with a test
experiment done simultaneously with the photon detection.

At low temperature, the molecules are described by the Hamiltonian%
\[
\mathcal{H}=-DS_{z}^{2}+g{\mu _{B}}S_{z}H+\mathcal{H}^{\prime }
\]%
where $S=10$ is the spin, $D=0.292K$ is the anisotropy parameter, $H$ is the applied magnetic field, $\mu _{B}$ is
the Bohr magneton, $g\approx 2$ is the gyromagnetic factor, and $\mathcal{H}%
^{\prime }$ does not commute with $S_{z}$ and is responsible for tunneling
between spin projection states $m$ \cite{barra1996,Caciuffo1998}. When the
field is strong ($\sim \mp 1$~T) only the $m=\pm S$ are populated. When the
field is swept across zero and changes sign, the $m=\pm S$ state becomes
metastable. At matching fields, which are separated by $0.225$~T \cite%
{Wernsdorfer2000,Caneschi1999}, quantum tunneling of magnetization (QTM) can
take place from $m=\pm S$ to an $m^{\prime }=\mp (S-n)$ state, where $%
n=0,1,2\ldots $ is an excited state index. For $n>0$, the excited $m^{\prime
}$ state decays spontaneously to the ground state $\mp S$ and energy is
emitted. For $n=1$ this energy corresponds to a frequency of $115.6$~GHz or
wavelength of $2.6$~mm and for $n=2$ it corresponds to a frequency of $219$~GHz. Interestingly, we found (see below) that the heat is
released for a longer time from the second excited ($n=2$) state than from
the first one ($n=1$).

The experiment is preformed below $0.2$~K in order to have temperature
independent quantum tunneling \cite{Wernsdorfer2000,Caneschi1999}. Fig.~\ref%
{fig1} depicts the experimental setup, which is located inside the inner
vacuum chamber of the DR. The cooling of the sample and all detectors is provided
via copper cold fingers attached to the mixing chamber (MC) of the DR. The
magnetization is measured using a Hall sensor array placed at the center of
a magnet. The array is made of Hall bars of dimensions 100$\times $100 $\mu $%
m$^{2}$ with 100 $\mu $m interval; the active layer in these sensors is a
two-dimensional electron gas formed at the interface of GaAs/AlGaAs
heterostructure. The surface of the Hall sensor is parallel to the applied
field. Consequently, the effect of the applied field on the sensor is
minimal and determined only by the ability to align the array surface and
field. The sample with the Hall sensor is located in the middle of a
copper cylinder, which acts as a wave guide and is also thermally linked to the
MC.

\begin{figure}[h]
\begin{center}
\includegraphics[width=\columnwidth]{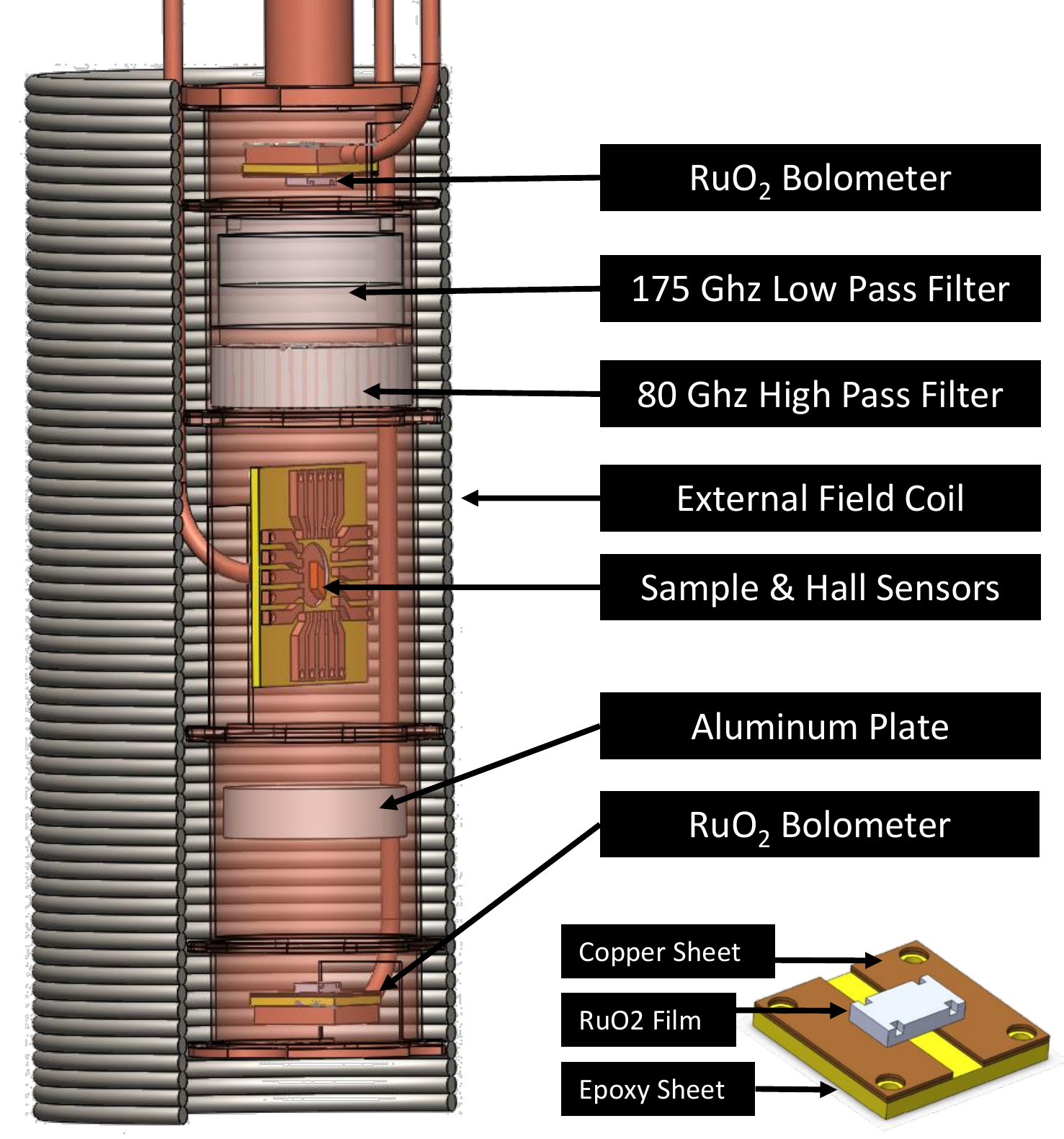}\\[0pt]
\end{center}
\caption{Experimental Setup showing the Hall sensors responsible for the
magnetization measurements, the bolometers with filters which detect
photons, the magnet, and the cold fingers linked to the mixing chamber.}
\label{fig1}
\end{figure}

Two bolometers are located at both ends of the cylinder. The bolometer
configuration is also shown in Fig.\ref{fig1}. The bolometers are made of a
RuO$_{2}$ thermistor attached to absorbing sheets. The thermistor is a
standard LakeShore RX-202A with typical temperature dependent resistance. The
thermistor is soldered from both sides to the copper sheets and copper
GE-varnish coated wires are soldered to the sheets.
The thermistors are biased by AC current of $10$ nA and their voltage is measured with a lock-in amplifier.

The absorbing sheets consisted of two copper plates $11$mm$%
\times 4$mm$\times 35\mu $m in size, with a gap between them. The
RuO$_{2}$ thermistor is bridging the gap. A thermally isolating layer of Glass Epoxy FR-4 is placed under the absorbing sheets.
The bolometers are mounted on a printed circuit board and have a weak thermal link to the MC.


Between one of the bolometers and the sample there is a combination of two filters making
a 80-180 Ghz band pass. The high pass is a `thick grill filter' based on
waveguide cut-off  \cite{Timusk1981} and the low pass is based on a mesh grid \cite{Ulrich1967}. We will refer to this bolometer and filter combination as the \textquotedblleft open\textquotedblright\ side. The other bolometer is totally blocked from radiation by a thick aluminum plate. The blocked side serves as the test experiment; the radiation is to be detected by the open bolometer only.  The band-pass filter was tested at room
temperature, using Spacek Labs GW-110-10 Gunn oscillator source operating
at 110~GHz and DW-2P broad band detector.

We test the response of the bolometers to a pulse of radiation in-situ
by replacing the Fe$_8$ sample with two Fairchild LED56 diodes that are pointing in both directions of the cylinder. The diodes are thermally connected directly to the $1$K pot of the DR for better
cooling power. The diodes' bias power is selected so as to give
a similar energy pulse to the bolometers as a tunneling event with the Fe$_{8}$
sample (see below). In Fig.~\ref{fig2} we plot the open and blocked
bolometers voltage as a function of time after energizing the diodes. The
solid line indicates the voltage across the diodes as a function of time.
The bolometer voltage is proportional to the temperature of the thermistor.
The temperature of the bolometer which is open to radiation increases more
and earlier than the blocked one. A few seconds later, the thermal energy from the diodes reaches both bolometers simultaneously. We also test the ability of the two bolometers
to detect thermal energy. The inset of Fig.~\ref{fig2} shows the case when the sample
area is heated by a resistor. The power and duration of this heat pulse
are again similar to that produced by the Fe$_{8}$ sample (see below). In
this case the temperature of both bolometers increases simultaneously to equal temperature.
Therefore, by subtracting the voltage of the bolometers, and
focusing on the early time before thermal energy arrives to the bolometers, we obtain the
signal of electromagnetic radiation only. This signal is also depicted in Fig.~\ref{fig2} and
demonstrates that we can clearly detect electromagnetic radiation emitted
from the diodes using our experimental setup.

\begin{figure}[h]
\begin{center}
\includegraphics[width=\columnwidth]{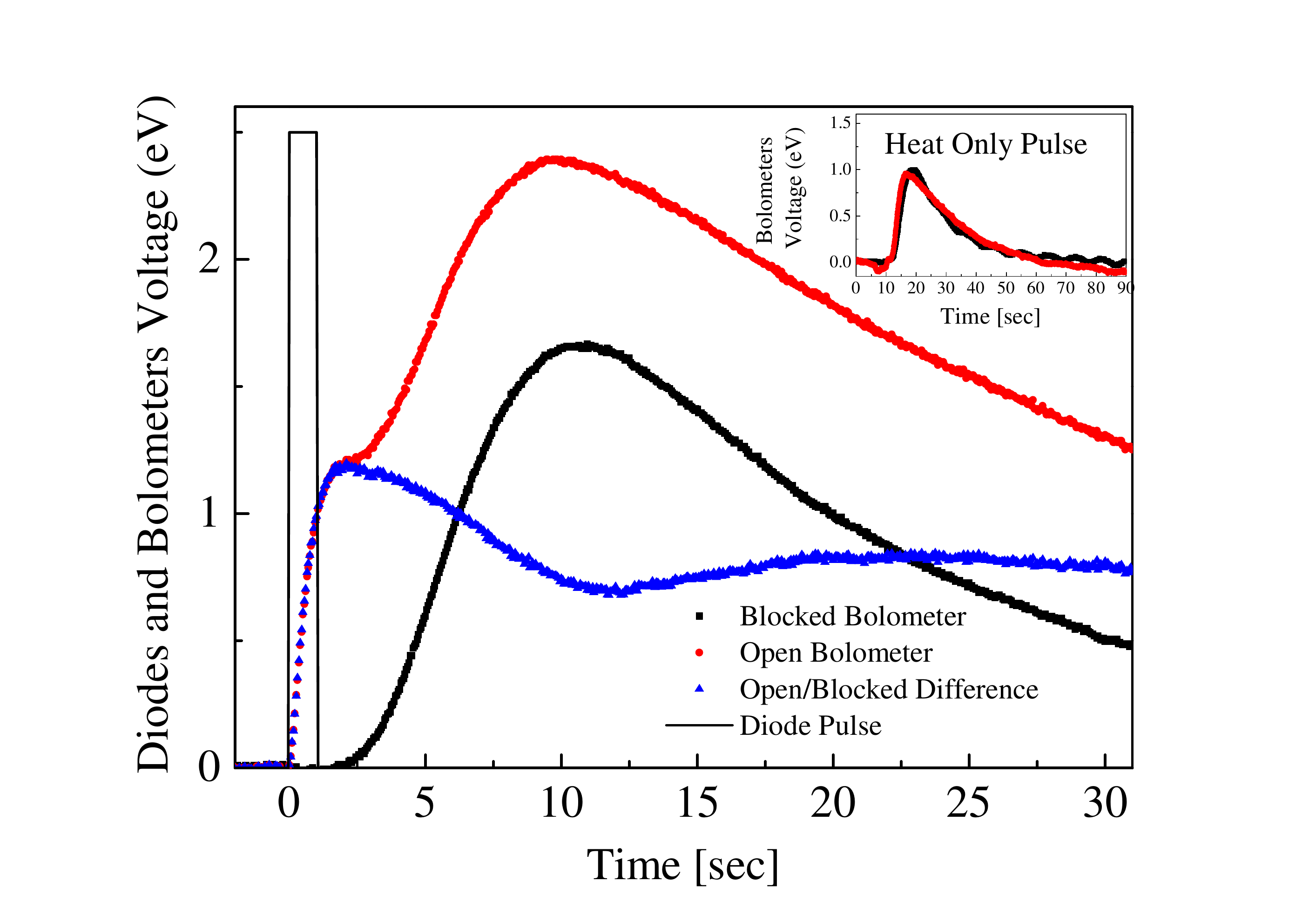}\\[0pt]
\end{center}
\caption{The response of the bolometers to a test radiation and thermal energy pulses.
The solid line is the voltage pulse applied to the light emitting diodes. The red and green symbols show the voltage developed across the open and blocked bolometers as their temperature increases due to the radiative pulse from the diodes. The blue symbols is the voltage difference. The difference within the first second represents a detection of a photon signal.
The inset shows the same experiment but with an input of thermal energy into the sample using a
biased resistor.}
\label{fig2}
\end{figure}

\begin{figure}[h]
\begin{center}
\includegraphics[width=\columnwidth]{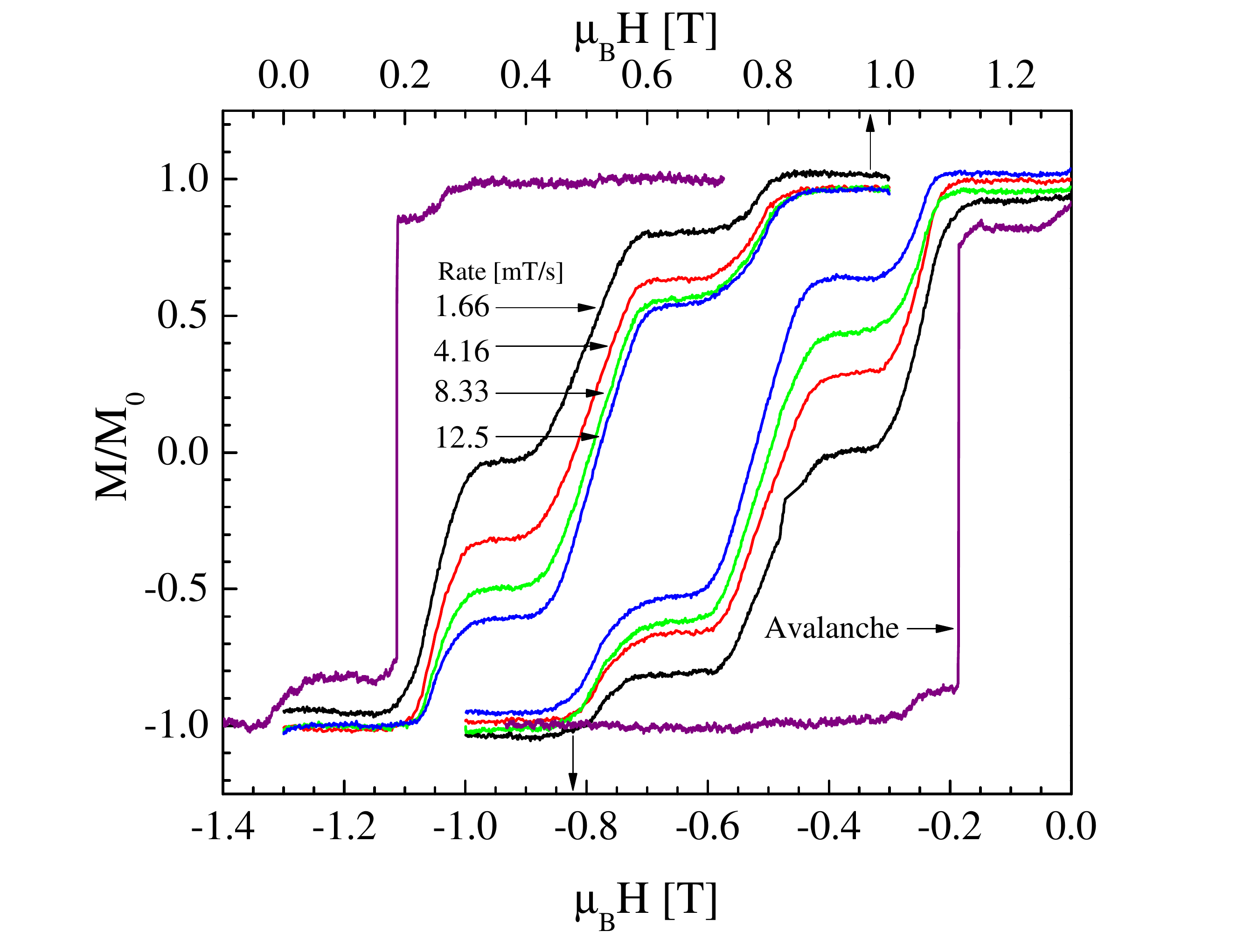}\\[0pt]
\end{center}
\caption{Fe$_{8}$ hysteresis loops with multi-step
magnetization jumps, at different sweep rates, and a hysteresis loop with an avalanche.
The fields for the positive sweep rates are given by the bottom abscissa, and for the negative sweep rates by the top abscissa.}
\label{fig3}
\end{figure}

In the experiment with Fe$_{8}$, the molecules are polarized by applying a
magnetic field of $\pm 1$~T in the $\mathbf{\hat{z}}$ direction. Afterwards,
the magnetic field is swept to $\mp 1$~T. The sweep is done at different
sweep rates. During the sweep we record the Hall voltages, the external
field, and the bolometers' voltage. The normalized magnetization $M/M_{0}$ is
given by the Hall voltage divided by the voltage at $H=1$~T. We found that
depending on the sweep rate, magnetization reversal can occur in two
different ways: a continuous reversal with multiple steps at matching
fields, or fast abrupt reversal, in avalanche form, as shown in Fig. \ref{fig3}.
We look for electromagnetic radiation in both cases. In the avalanche process, a large amount of heat is released and a clear tunneling front is present \cite{Leviant1}. Without
avalanche, the temperature of the sample remains low compared to the energy
barrier. In this case, a unique quanta of energy should be emitted in the
tunneling process.

\begin{figure}[H]
\begin{center}
\includegraphics[width=\columnwidth]{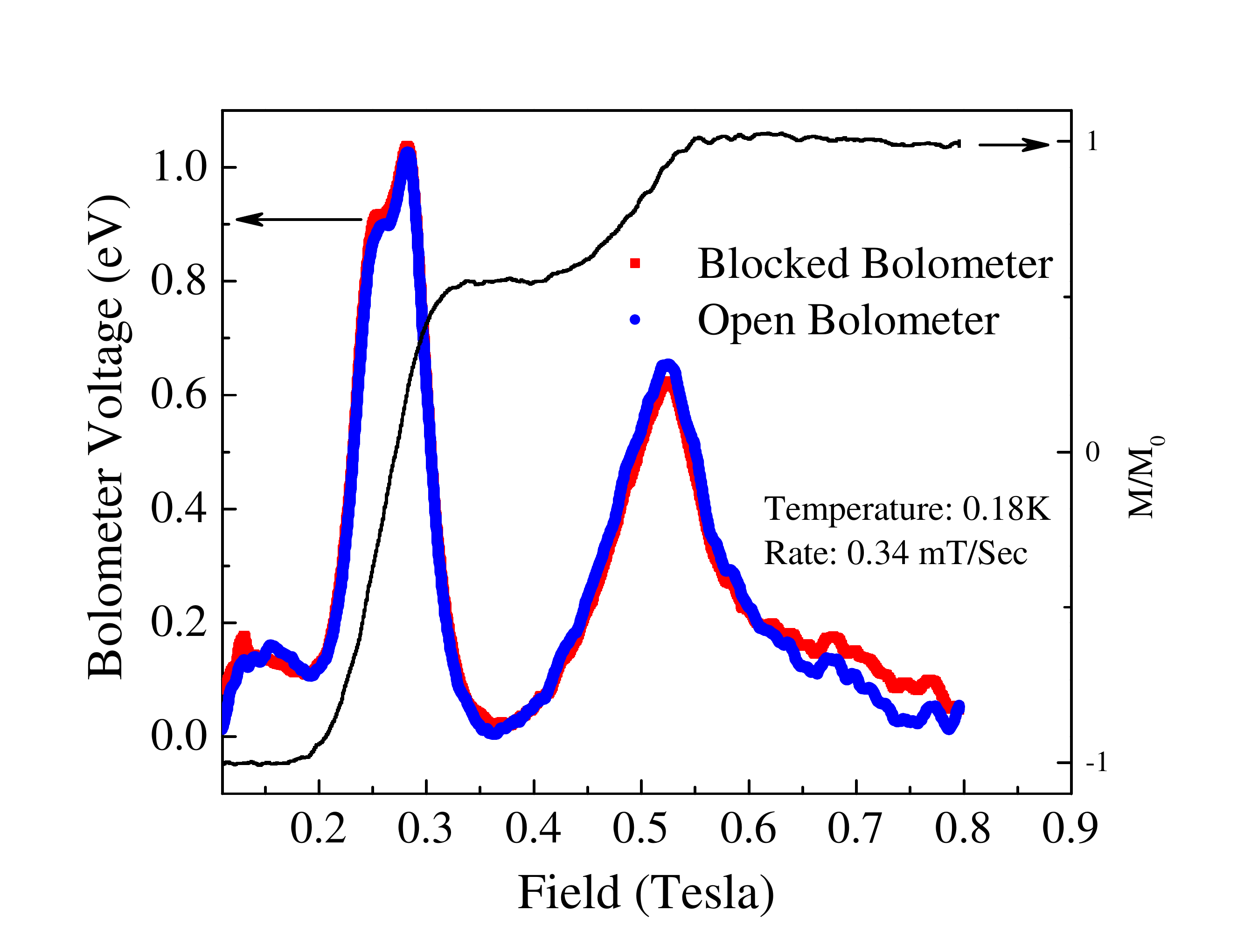}\\[0pt]
\end{center}
\caption{Magnetization and energy emission measurements done simultaneously
on an Fe$_{8}$ molecular magnet. The left ordinate is the open and closed
bolometers voltage, which is proportional to their temperature. The right
ordinate is the normalized magnetization. No difference between the two
bolometers is detected within the experimental sensitivity. The second
bolometers voltage peak decays more slowly than the first one, with no
noticeable magnetization changes at fields approaching $1$~T.}
\label{fig4}
\end{figure}

The results of our experiment in the case of multiple magnetization steps
are shown in Fig.~\ref{fig4}. The left ordinate is the bolometers voltage.
The right ordinate is the normalized magnetization $M/M_{0}$. Both are
plotted as a function of applied magnetic field. The external field is swept
at a rate of $0.34$~mT/sec from positive to negative.
When the external field is at matching value a QTM occurs followed by rise of the bolometers
voltage/temperature. However, there is no difference between the opened and
the blocked bolometer for neither of the transitions. The same results are
obtained at higher sweep rates with avalanches. Therefore, we can not find
indication of electromagnetic radiation emanating from Fe$_{8}$ regardless
of the sweep rate or transition index $n$. This is the main finding of this
work.

However, it is interesting to notice that at the second transition it takes
the bolometers more time (longer field interval) to cool down than at the
first transition. This could have two possible explanations: (I) The lifetime
of the $n=2$ excited state is longer than the $n=1$ state. This
possibility stands in contrast to lifetime measurements by Bahr et al. \cite%
{Bahr2008}, although they where done at higher temperatures. (II) As we
sweep the field there are more transitions from the metastable state to $n=3,
$ $4\ldots $ excited states. As $n$ increases the magnetization change
becomes smaller but the energy released becomes larger. It is conceivable that we
are unable to detect magnetically the higher transitions but can detect
their energy release. More experiments are required to distinguish between
the two possibilities.

To summarize, we re-examine the possibility that Fe$_{8}$ emits
electromagnetic radiation after tunneling events using a specially designed
experimental setup. With this experiment we can not detect photon emission
and can not reproduce the results that were reported previously  \cite{Oren}.
We conclude that energy is released after tunneling in
Fe$_{8}$ only in the form of thermal energy. This is important for understanding the role
of phonons in the tunneling process.

The study was partially supported by the Norman and Helen Asher foundation
for space research and by the Russell Berrie Nanotechnology Institute,
Technion, Israel Institute of Technology

\end{document}